\documentclass[letterpaper, 10 pt, conference]{ieeeconf}  
\usepackage{graphicx}
\usepackage{hyperref}
\usepackage[margin=0.75in]{geometry}
\makeatletter
\newcommand*{\rom}[1]{\expandafter\@slowromancap\romannumeral #1@}
\makeatother

\IEEEoverridecommandlockouts                              

\title{\LARGE \bf
Brain Signals to Rescue Aphasia, Apraxia and Dysarthria Speech Recognition 
}

\author{Gautam Krishna$^{1}$, Mason Carnahan$^{1}$, Shilpa Shamapant$^{2}$, Yashitha Surendranath$^{1,*}$, Saumya Jain$^{1,*}$,\\
Arundhati Ghosh$^{1,*}$,Co Tran$^{1}$, Jose del R 
Millan$^{3}$ and Ahmed H Tewfik$^{1}$
\thanks{$^{1}$Author is with Brain Machine Interface Lab, The University of Texas at Austin,
        Austin, Texas, USA
        {}}%
\thanks{$^{2}$Author is with Austin Speech Labs, Austin,
        Texas, USA
        {}}%
\thanks{$^{3}$Author is with Clinical Neuroprosthetics and Brain Interaction Lab, The University of Texas at Austin,
        Austin, Texas, USA
        {}}%
\thanks{$^{*}$Equal author contribution 
        {}}%
}

\begin{document}

\maketitle
\thispagestyle{empty}
\pagestyle{empty}

\begin{abstract}

In this paper, we propose a deep learning-based algorithm to improve the performance of automatic speech recognition (ASR) systems for aphasia, apraxia, and dysarthria speech by utilizing electroencephalography (EEG) features recorded synchronously with aphasia, apraxia, and dysarthria speech. We demonstrate a significant decoding performance improvement by more than 50\% during test time for isolated speech recognition task and we also provide preliminary results indicating performance improvement for the more challenging continuous speech recognition task by utilizing EEG features. The results presented in this paper show the first step towards demonstrating the possibility of utilizing non-invasive neural signals to design a real-time robust speech prosthetic for stroke survivors recovering from aphasia, apraxia, and dysarthria. Our aphasia, apraxia, and dysarthria speech-EEG data set will be released to the public to help further advance this interesting and crucial research. 

\end{abstract}

\section{INTRODUCTION}
Automatic speech recognition (ASR) system converts speech to text and it forms the back-end in many state-of-the-art virtual voice assistants like Apple's Siri, Amazon's Alexa, Samsung's Bixby, etc. These voice assistants are trained to recognize the uniform speech of users with no speech disorders. The performance of ASR systems degrades in presence of incomplete, distorted, or broken speech. This limits technology accessibility to users with speech disorders. The three most common speech, language disorders are aphasia, apraxia, and dysarthria. Aphasia is a disturbance of the comprehension and formulation of language caused by dysfunction in specific brain regions.  The major causes are a stroke or head trauma\cite{damasio1992aphasia,benson1996aphasia}. Apraxia is a speech disorder caused due to the impairment of motor planning of speech \cite{kent1983acoustic}. Dysarthria is also a speech disorder caused due to neurological damage to the motor component of the motor–speech system and it is closely related to Apraxia \cite{darley1969differential}. People recovering from these speech disorders produce distorted and incomplete speech.  The work described by authors in \cite{krishna2020synthesis,anumanchipalli2019speech} demonstrate that electrophysiological monitoring of neural signals like electroencephalography (EEG) and electrocorticography (ECoG) carry important information about speech articulation and speech perception. They demonstrated the results using neural signals recorded from subjects with no speech disorders. In \cite{krishna2019speech} authors demonstrated that EEG features can be used to enhance the performance of isolated speech recognition systems trained to decode speech of users with no speech disorders. In their work, they demonstrated results on an English vocabulary consisting of four words and five vowels. EEG is a non-invasive way of measuring the electrical activity of the human brain. The EEG sensors are placed on the scalp of the subject to obtain EEG recordings. The EEG signals offer a very high temporal resolution. The non-invasive nature of EEG signals makes it safe and easy to deploy even-though EEG signals offer poor spatial resolution and signal-to-noise ratio compared to invasive ECoG neural activity recording techniques. The high temporal resolution property of EEG signals also allows capturing the human speech-related neural activities as normal human speech occurs at a high rate of 150 words per minute. 
In \cite{fraser2013automatic} authors explored speech recognition using aphasia speech and reported a very high word error rate (WER) during test time. For a reduced vocabulary, they reported a WER as high as 97.5 \%. In \cite{le2016improving} authors demonstrated aphasia speech recognition by training their acoustic models on a large scale aphasia speech data-set named AphasiaBank but they reported a high phoneme error rate (PER) in the range of 75\% to 89\% for severe cases of aphasia. A high PER indicates an even higher WER. In a very recent work described in \cite{ballard2019feasibility} authors explored the possibility of using ASR systems as a feedback tool while providing speech therapy to aphasia patients. Their results demonstrated an increase in the effectiveness of the speech therapy when coupled with ASR technology. References \cite{jacks2019automated,green2003automatic,ferrier1995dysarthric} investigated speech recognition for apraxia and dysarthria speech and reported low accuracy on a word-level vocabulary. 
In \cite{spironelli2009eeg} authors carried out an EEG study to analyze the EEG delta wavebands to understand the brain damage on patients recovering from aphasia. In related studies described in references \cite{hensel2004left,sarasso2014plastic} authors investigated EEG activity in the left-hemisphere of the brain of subjects recovering from aphasia and an EEG sleep study to understand the brain activity of the aphasia patients. These studies demonstrated that EEG signals carried useful information about brain function recovery in aphasia patients. 
In this paper, we propose an algorithm to train a deep learning-based speech recognizer using acoustic features along with acoustic representations derived from EEG features to significantly improve the test time decoding performance of aphasia + apraxia + dysarthria isolated speech recognizer. We were able to achieve a performance improvement of more than 50\% during test time for the task of isolated speech recognition and a slight improvement in performance for the more challenging task of continuous speech recognition using our proposed algorithm. The results presented in this paper demonstrate how non-invasive neural signals can be utilized to improve the performance of speech recognizers used to decode aphasia, apraxia, and dysarthria speech. Designing a speech recognizer that can decode aphasia, apraxia, and dysarthria speech with high accuracy has the potentiality to lead to a design of a speech prosthetic and a better speech therapy tool for stroke survivors.

Our main contributions and major highlights of our proposed algorithm are listed below:
\begin{itemize}
    \item We developed a deep learning-based algorithm to improve the performance of speech recognition for aphasia, apraxia, and dysarthria speech by utilizing EEG features. 
    \item We collected large-scale aphasia, apraxia and dysarthria Speech-EEG data set that will be released to the public to help further advance this research.  
    \item Our proposed algorithm can be used with any type of speech recognition model, for example in this work we demonstrate the application of the algorithm on isolated as well as continuous speech recognition models. 
\end{itemize}

\section{Proposed Deep learning algorithm to improve Speech Recognition}
Figure 1 describes the architecture of our proposed deep learning training strategy to improve the ASR performance of aphasia, apraxia, and dysarthria speech by utilizing EEG features. As seen from the figure, we make use of an EEG to acoustic feature mapping, regression model to generate additional features that are provided to the ASR model to improve its training. We first train the regression model described on the right-hand side of the figure to predict acoustic features or Mel frequency cepstral coefficients (MFCC) \cite{vergin1999generalized} of dimension 13 from EEG features. The regression model consists of a single layer of gated recurrent unit (GRU) \cite{chung2014empirical} with 128 hidden units connected to a time distributed dense layer consisting of 13 hidden units with a linear activation function. The regression model was trained for 70 epochs with mean square error (MSE) as the loss function and with adam \cite{kingma2014adam} as the optimizer. The batch size was set to 100. The GRU layer in the regression model learns the acoustic representation present in the input EEG features. We then concatenate these acoustic representations or outputs of the GRU layer of the regression model with the input acoustic or MFCC features of dimension 13 which are then used to train the ASR model to produce the text output during training time. The ASR model is trained after completing the training of the regression model. During test time, the EEG features from the test set are provided as input to the trained regression model, and the output of the GRU layer of the regression model is concatenated with the simultaneously recorded MFCC features from the test set to produce text output from the trained ASR model.  The output of the GRU layer of the regression model or the acoustic representations present in EEG features is of dimension 128. The choice of the ASR model architecture depends on the task. We investigated both the tasks of isolated and continuous speech recognition in this paper. Isolated speech recognition refers to a sentence or sequence classification task, where the model decodes closed vocabulary and directly learns the input feature to sentence mapping. Here the model predicts the complete sentence or label token as output per decoding stage. On the other hand, continuous speech recognition refers to the task where the model is predicting the text by predicting the character or word or phoneme at every time-step and these models are capable of performing open vocabulary decoding. Continuous speech recognition is a more challenging task than isolated speech recognition due to the increase in the number of model parameters and learning alignments. 
\begin{figure}[h]
\begin{center}
\includegraphics[height=7.0cm, width=0.5\textwidth,trim={0.1cm 0.1cm 0.1cm 0.1cm}]{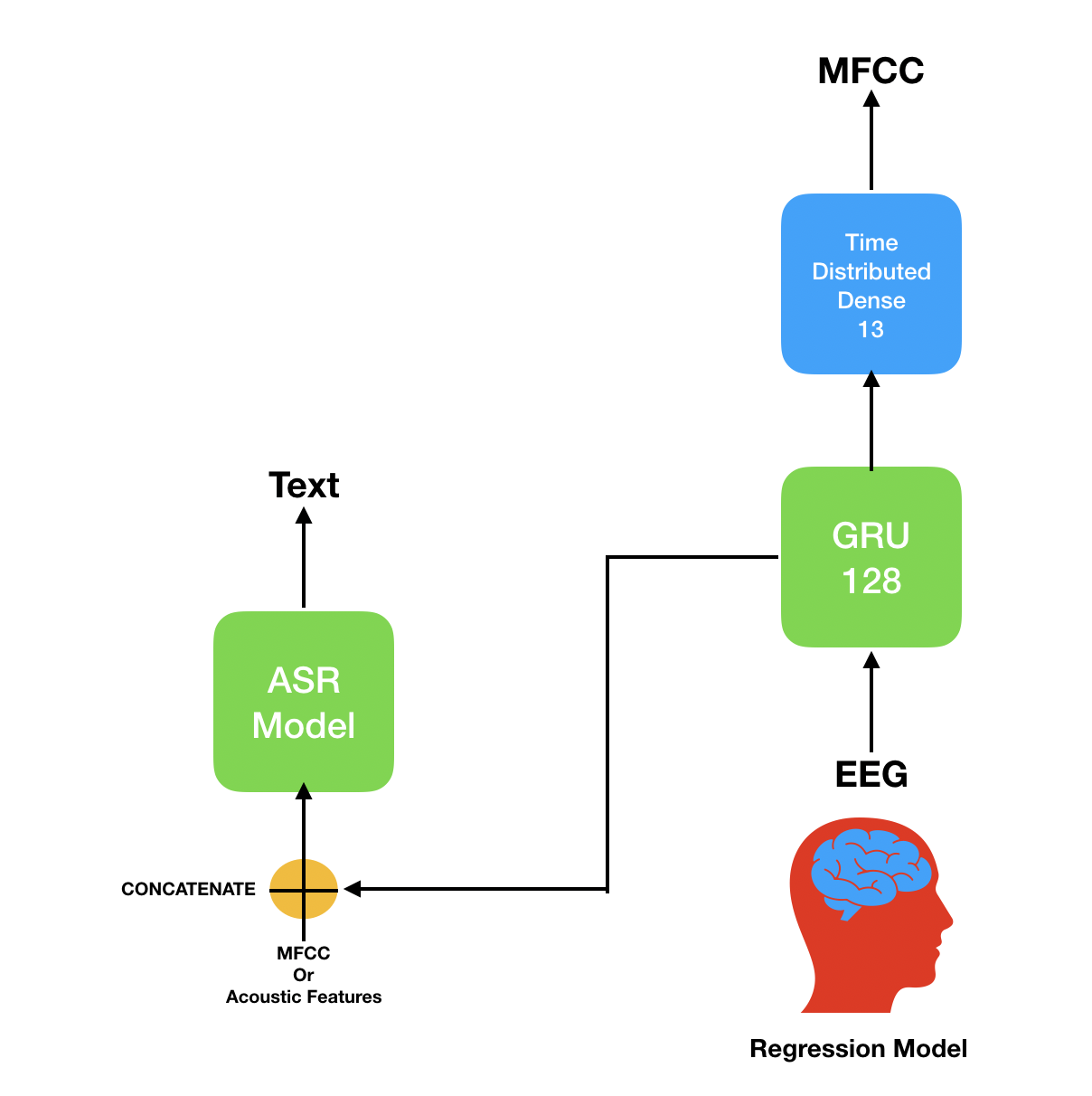}
\caption{Proposed Training Algorithm} 
\label{1vsall}
\end{center}
\end{figure}

\begin{figure}[h]
\begin{center}
\includegraphics[height=7.0cm, width=0.4\linewidth,trim={0.1cm 0.1cm 0.1cm 0.1cm}]{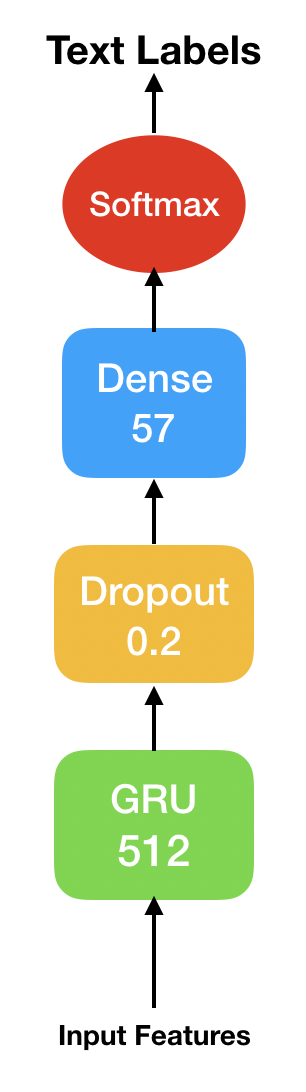}
\caption{Isolated Speech Recognition Model} 
\label{1vsall}
\end{center}
\end{figure}

Next, we briefly describe the architecture of the isolated and continuous speech recognition models used in this work. Our isolated speech recognition model consists of a single layer of GRU with 512 hidden units connected to a dropout regularization \cite{srivastava2014dropout} with a drop-out rate of 0.2. The drop-out regularization is followed by a dense layer consisting of 57 hidden units and a linear activation function. The dense layer contained 57 hidden units since our vocabulary contained 57 unique English sentences. The last time-step output of the GRU layer is passed to dropout regularization and dense layer. Finally, the dense layer output or logits are passed through a softmax activation function to obtain the label prediction probabilities. Each label token corresponds to a complete English sentence text. The labels were one-hot vector encoded and the model was trained for 10 epochs with batch size set to 50. We used early stopping during training to prevent over-fitting. We used categorical cross-entropy as the loss function and adam was used as the optimizer. The model architecture is described in Figure 2. 
Our continuous speech recognition model consists of a GRU layer with 512 hidden units acting as an encoder and the decoder consists of a combination of a dense layer with linear activation function and softmax activation function. The output of the encoder is passed to the decoder at every time-step.  The model was trained for 100 epochs with batch size set to 50 to optimize connectionist temporal classification (CTC) loss function \cite{graves2006connectionist,graves2014towards}. We used adam as the optimizer. For this work, a character-based CTC model was used. The model was predicting a character at every time-step. We used an external 4-gram language model along with a CTC beam search decoder during inference time \cite{toshniwal2018comparison}. 

\begin{figure}[h]
\begin{center}
\includegraphics[height=7.0cm, width=0.4\linewidth,trim={0.1cm 0.1cm 0.1cm 0.1cm}]{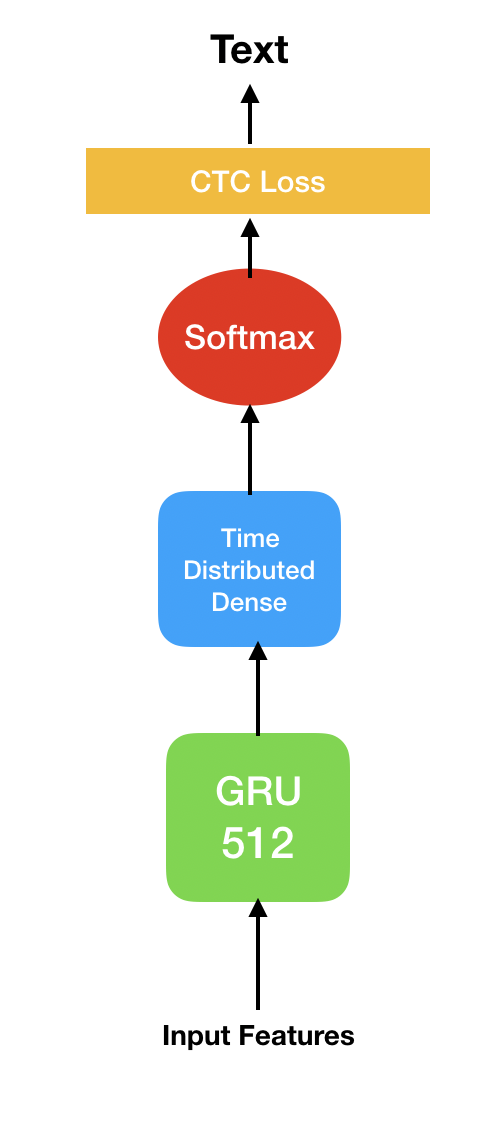}
\caption{Continuous Speech Recognition Model} 
\label{1vsall}
\end{center}
\end{figure}

\section{Design of experiments for building the data base}

Nine subjects with a known diagnosis of aphasia or apraxia or dysarthria or a combination of any of these disorders volunteered to take part in our data collection experiments. All experimental procedures were approved by the Institutional Review Board at the University of Texas at Austin. The demographic information of the subjects is shown below in Table \rom{1}. Each subject was asked to perform two different tasks while they were receiving speech therapy at Austin Speech Labs. The first task involved subjects reading out loud English sentences shown to them on a computer screen and their simultaneous EEG, electromyography (EMG), and speech signals were recorded. The second task involved subjects listening to the recorded audio of English sentences and they were then asked to speak out loud what they listened to and their simultaneous EEG, EMG, and speech signals were recorded. We collected a total of 8854 data samples from the 9 subjects for both the tasks combined. The vocabulary consisted of 57 unique daily used common English sentences.  We used brain products wet EEG sensors for this data collection. We used 29 EEG sensors in total. The placement of 29 sensors was based on the standard 10-20 EEG sensor placement guidelines. Figure 4 shows a subject wearing our EEG cap during the experiment. We used the brain product's Actchamp amplifier as the EEG amplifier. 
We further used two EMG sensors to keep track of EMG artifacts generated during articulation. The EMG sensor placement location is shown in Figure 5. The speech signals were recorded using a mono-channel microphone. We used 70\% of the data as the training set, 10\% as the validation set, and the remaining 20\% as the test set. The data set split was done randomly using the scikit-learn train-test split python function. There was no overlap between training, validation, and test set data points. 

\begin{table}[!ht]
\tiny
\centering
\begin{tabular}{|l|l|l|l|l|l|}
\hline
\textbf{ID} & \textbf{\begin{tabular}[c]{@{}l@{}}Male/\\ Female\end{tabular}} & \textbf{\begin{tabular}[c]{@{}l@{}}Aphasia\\ Quotient\end{tabular}} & \textbf{\begin{tabular}[c]{@{}l@{}}Aphasia\\ Type\end{tabular}} & \textbf{Severity} & \textbf{Speech Disorders} \\ \hline
1           & M                                                               & 48.4                                                                & global                                                          & severe            & aphasia, apraxia          \\ \hline
2           & F                                                               & 53.2                                                                & global                                                          & severe            & aphasia                   \\ \hline
3           & M                                                               & 74.2                                                                & Broca's                                                         & moderate          & aphasia                   \\ \hline
4           & M                                                               & 87.6                                                                & Anomia                                                          & mild              & aphasia                   \\ \hline
5           & M                                                               & 86.2                                                                & Broca's                                                         & moderate          & aphasia, apraxia          \\ \hline
6           & M                                                               & 94.8                                                                & Anomic                                                          & mild              & aphasia, dysarthria       \\ \hline
7           & M                                                               & 48                                                                  & global                                                          & severe            & aphasia                   \\ \hline
8           & M                                                               & 87.4                                                                & Broca's                                                         & moderate          & aphasia, apraxia          \\ \hline
9           & M                                                               & 72.4                                                                & mixed                                                           & mild              & aphasia                   \\ \hline
\end{tabular}
\caption{Data-set Demographics}
\end{table}

\begin{figure}[h]
\begin{center}
\includegraphics[height=4.5cm, width=0.5\linewidth,trim={0.1cm 0.1cm 0.1cm 0.1cm}]{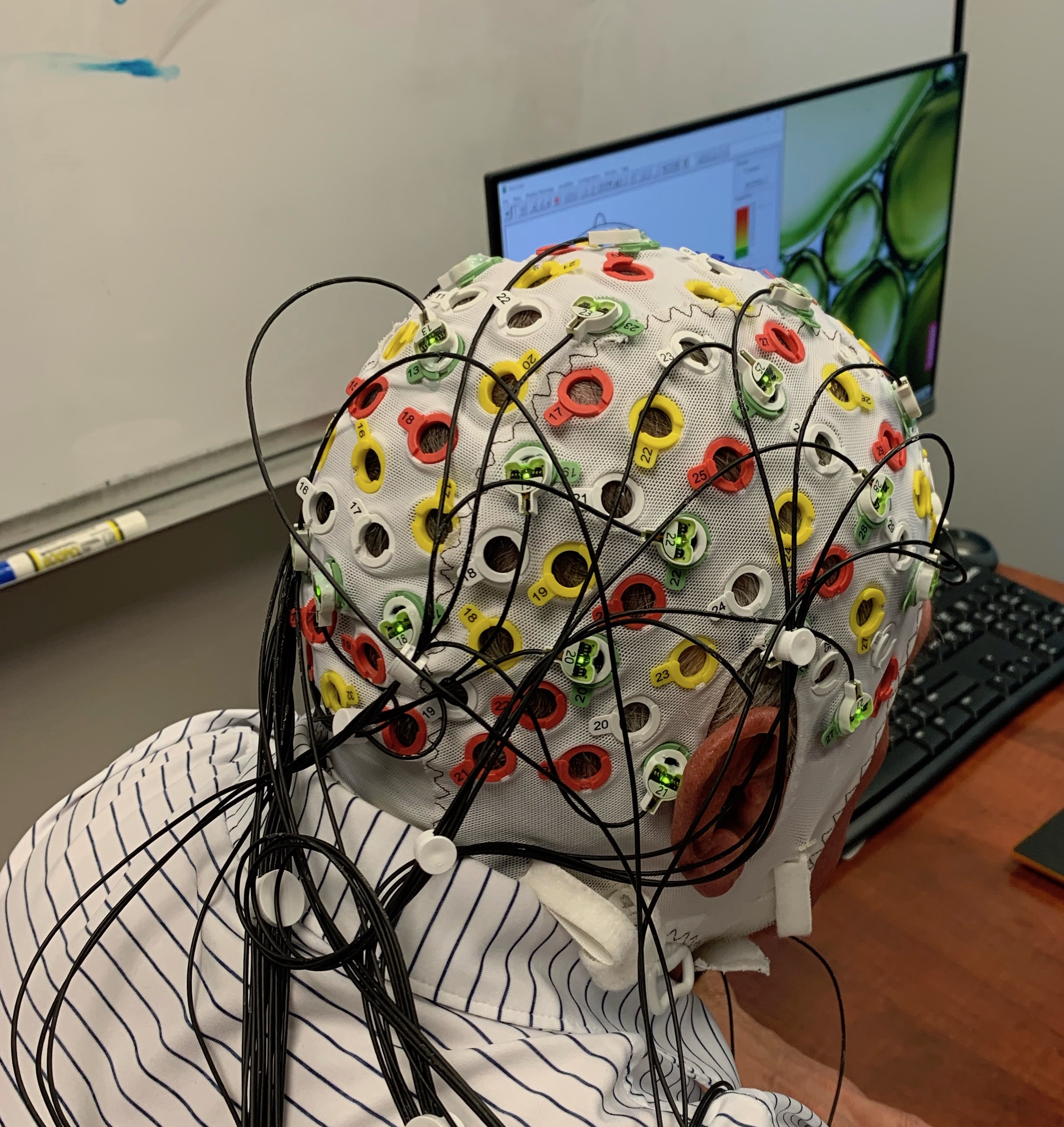}
\caption{EEG sensor placement} 
\label{1vsall}
\end{center}
\end{figure}

\begin{figure}[h]
\begin{center}
\includegraphics[height=4.5cm, width=0.6\linewidth,trim={0.1cm 0.1cm 0.1cm 0.1cm}]{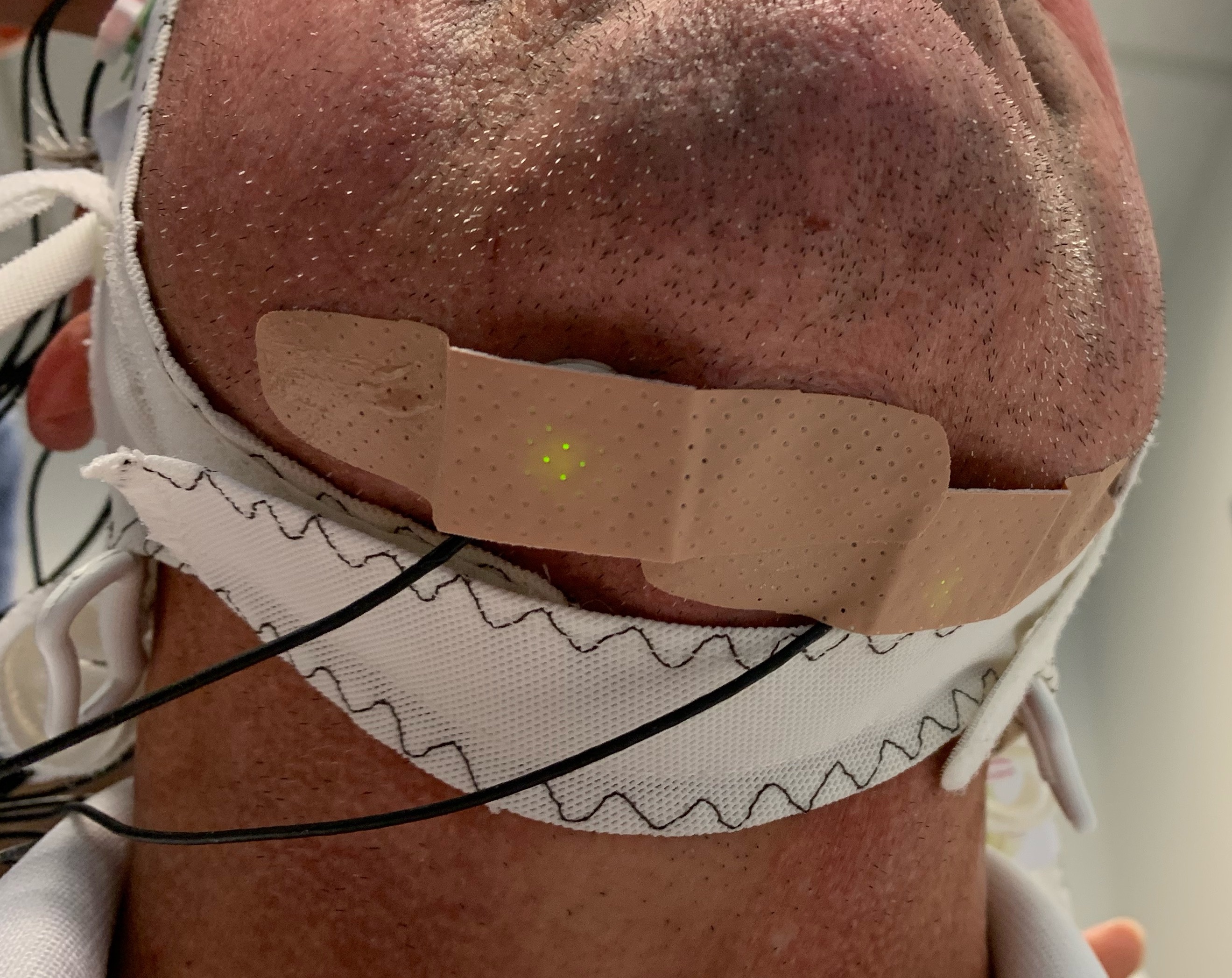}
\caption{EMG sensor placement} 
\label{1vsall}
\end{center}
\end{figure}

\section{EEG and Speech feature extraction details}
The recorded EEG signals were sampled at a sampling frequency of 1000Hz and a fourth-order IIR bandpass filter with cut-off frequencies 0.1Hz and 70Hz was applied. A notch filter with a cut off frequency of 60 Hz was used to remove the power line noise. We then used the linear regression technique to remove EMG artifacts from EEG signals. \\${Corrected}_{EEG}$ = ${Recorded}_{EEG} - \alpha \ast {Recorded}_{EMG} $, where $\alpha$ is the regression coefficient computed by Ordinary Least Squares method. 
We then extracted five features per EEG channel. The five features extracted were root mean square, zero-crossing rate, moving window average, kurtosis, and power spectral entropy \cite{krishna2019speech,krishna20}. This EEG feature set was first introduced by authors in \cite{krishna2019speech} where they demonstrated that these features carry neural information about speech perception and production. The EEG features were extracted at a sampling frequency of 100 Hz per channel. 
The speech signal was recorded at a sampling frequency of 16KHz. We extracted Mel frequency cepstral coefficients (MFCC) \cite{vergin1999generalized} of dimension 13 as features for speech signal. The MFCC features were also extracted at the same sampling frequency 100Hz as that of EEG feature extraction. 


\begin{figure}[h]
\begin{center}
\includegraphics[height=4.5cm, width=0.7\linewidth,trim={0.1cm 0.1cm 0.1cm 0.1cm}]{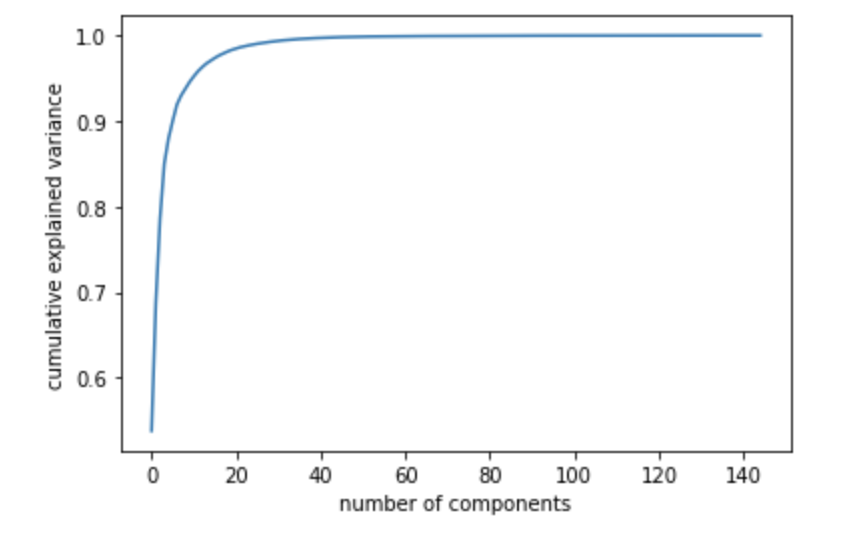}
\caption{Explained Variance vs No of Components} 
\label{1vsall}
\end{center}
\end{figure}







\section{EEG Feature Dimension Reduction Algorithm Details}

Since the local structure of our EEG feature space was not linear, we used non-linear dimension reduction technique to perform dimension reduction on EEG features. We plotted cumulative explained variance vs the number of components as shown in Figure 6 to identify the optimal EEG feature space dimension. We used kernel principal component analysis (KPCA) \cite{mika1999kernel} with a polynomial kernel of degree 3 to reduce our EEG feature space of dimension 145 (five features per each of the 29 channels) to a final dimension of 10. Before applying KPCA, the EEG features were normalized by removing the mean and scaling to unit variance. 

\section{Results and Discussion}

We used percentage accuracy, F1-score, precision, and recall \cite{goutte2005probabilistic} as performance metrics to evaluate the performance of the isolated speech recognition model. The higher the accuracy, F1-score, precision, and recall values the better the performance of the model. For computing F1-score, precision and recall we added a small value e-07 called  \href{https://www.tensorflow.org/api_docs/python/tf/keras/backend/epsilon}{epsilon} to the denominator of F1-score, precision and recall expressions to prevent a divide by zero error. 
We used word error rate (WER) as the performance metric to evaluate the performance of the continuous speech recognition model. The lower the WER value, the better the speech recognition system performance. For obtaining baseline results, the speech recognition models were trained and tested using only acoustic or MFCC features. Table \rom{2} shows the test time results obtained for isolated speech recognition task for various EEG frequency bands. We compared results obtained using low-frequency EEG signals ( 0.1 Hz to 15 Hz), high-frequency EEG signals ( 15 Hz to 70 Hz), and all frequency EEG signals ( 0.1 Hz to 70 Hz). The results shown in Table \rom{2} demonstrate that choice of EEG frequency range had less effect on decoding performance for the isolated speech recognition task. The work carried out by authors in \cite{anumanchipalli2019speech} demonstrated that both high and low-frequency neural signals carry important information about speech production. 
Table \rom{3} shows test times results for isolated speech recognition task with and without EMG artifact removal and obtained results demonstrate that even though removing EMG artifacts improved the test-time performance of the model, the improvement was not that significant. Table \rom{4} shows the test time results for isolated speech recognition task with and without EEG dimension reduction. The results demonstrate that EEG dimension reduction using KPCA had resulted in significant performance improvement of the model during test time. 
Table \rom{5} shows the test time results for isolated speech recognition task when we used only temporal lobe EEG sensor features, frontal lobe EEG sensor features, and concatenation of temporal and frontal lobe EEG sensor features. The temporal and frontal lobe contains brain regions responsible for speech perception and production \cite{flinker2015redefining,chartier2018encoding}. EEG features from frontal lobe sensors Fp1, Fz, F3, F7, FT9, FC5, FT10 ,  FC6 , FC2 , F4 , F8 and Fp2 were extracted and then reduced to a dimension of 10 using KPCA. Similarly, EEG features were extracted from temporal lobe sensors T7, TP9, TP10, and T8 and then reduced to a dimension of 10 using KPCA. The results shown in Table \rom{5} demonstrate that it is possible to achieve comparable decoding performance for isolated speech recognition task using EEG sensors from just frontal and temporal lobe regions instead of using all the EEG sensors. Table \rom{6} shows the test time results for isolated speech recognition task when we used only the first half-length of the input EEG and MFCC features instead of the complete length of EEG or MFCC features for decoding text. The motivation here was to see if the model can decode text if we provide incomplete input as most of the aphasia or apraxia speech involves a lot of pauses in between.  As seen from the Table \rom{6} results we observed that when half the length of the input signal is used, the baseline results improved significantly but adding acoustic representation in EEG to MFCC features still outperformed the baseline for all the test-time performance metrics. We believe the baseline results improved when shorter sequences were used as input signal due to the fact that GRU can process shorter sequences more efficiently than longer input sequences \cite{chung2014empirical,bai2018empirical}. The overall results from Tables \rom{2},\rom{3},\rom{5} and \rom{6} show that adding acoustic representation in EEG features with MFCC features significantly outperform the baseline for all the test-time performance metrics for the task of isolated speech recognition using aphasia, apraxia, and dysarthria speech. Figure 7 shows the training and validation loss convergence for the regression model and Figure 8 shows the training and validation accuracy of the isolated speech recognition model. The training, validation loss values were comparable as well as the training and validation accuracy values, indicating the models didn't overfit. Figure 9 shows the confusion matrix obtained during test time for the isolated speech recognition task when tested using MFCC+ High-frequency EEG representation. Each token in the confusion matrix represents a complete English sentence from the test set. 
Table \rom{7} shows test time results for isolated speech recognition task when acoustic features were concatenated with acoustic representation in EMG features of dimension 10 compared to acoustic representations from EEG features of dimension 10. We extracted the same set of 5 features that we extracted for EEG for each EMG channel. The results show that the acoustic representations present in EMG is not significant compared to acoustic representation features present in EEG signals for boosting the performance of the speech recognizer. 
Table \rom{8} shows the test time average WER obtained for the continuous speech recognition task. The obtained results demonstrate that adding acoustic representation in High-frequency EEG features to MFCC outperformed the baseline for a test set vocabulary consisting of 1771 English sentences.  We obtained a p value \cite{griffiths2019statistical} of \textbf{0.0000213}, demonstrating high statistical significance for our result.  We further computed the test time WER's with 95 \% confidence level value and observed that for the baseline, the WER range was between 48.58\% and 51.1\% where as  using our proposed method, the WER range for the same confidence level value was between \textbf{44.25\%} and \textbf{47.13\%}. Therefore a thorough statistical analysis of our test time continuous speech recognition results demonstrate that our proposed method outperformed the baseline result. 

\begin{table*}[!ht]
\scriptsize
\centering
\begin{tabular}{|l|l|c|c|c|}
\hline
\textbf{\begin{tabular}[c]{@{}l@{}}Performance\\ Metric(\%)\end{tabular}} & \textbf{MFCC} & \textbf{\begin{tabular}[c]{@{}c@{}}MFCC\\ + \\ Acoustic Representation\\ in Low Freq EEG dim 10\end{tabular}} & \textbf{\begin{tabular}[c]{@{}c@{}}MFCC\\ +\\ Acoustic Representation\\ in High Freq EEG dim 10\end{tabular}} & \textbf{\begin{tabular}[c]{@{}c@{}}MFCC\\ +\\ Acoustic Representation\\ in All Freq EEG dim 10\end{tabular}} \\ \hline
\textbf{Accuracy}                                                         & 28.40         & 78.93                                                                                                 & \textbf{81.02}                                                                                        & 80.74                                                                                                \\ \hline
\textbf{F1 - score}                                                       & 34.93         & 81.65                                                                                                 & 82.86                                                                                                 & \textbf{83.25}                                                                                       \\ \hline
\textbf{precision}                                                        & 75.88         & 86.54                                                                                                 & 87.23                                                                                                 & \textbf{88.27}                                                                                       \\ \hline
\textbf{recall}                                                           & 23.06         & 77.40                                                                                                 & \textbf{79.02}                                                                                        & 78.91                                                                                                \\ \hline
\end{tabular}
\caption{Test time results for isolated speech recognition for various EEG frequency bands}
\end{table*}

\begin{table*}[!ht]
\tiny
\centering
\begin{tabular}{|l|l|c|c|}
\hline
\textbf{\begin{tabular}[c]{@{}l@{}}Performance\\ Metric(\%)\end{tabular}} & \textbf{MFCC} & \textbf{\begin{tabular}[c]{@{}c@{}}MFCC\\ +\\ Acoustic Representation\\ in All Freq EEG dim 10\\ \\  EMG artifacts\\ Removed\end{tabular}} & \textbf{\begin{tabular}[c]{@{}c@{}}MFCC\\ +\\ Acoustic Representation\\ in All Freq EEG dim 10\\ \\ No EMG artifacts\\ Removed\end{tabular}} \\ \hline
\textbf{Accuracy}                                                         & 28.40         & \textbf{80.74}                                                                                                                     & 80.46                                                                                                                                \\ \hline
\textbf{F1 - score}                                                       & 34.93         & \textbf{83.25}                                                                                                                     & 82.24                                                                                                                                \\ \hline
\textbf{precision}                                                        & 75.88         & \textbf{88.27}                                                                                                                     & 86.00                                                                                                                                \\ \hline
\textbf{recall}                                                           & 23.06         & \textbf{78.91}                                                                                                                     & \textbf{78.91}                                                                                                                       \\ \hline
\end{tabular}
\caption{Isolated Speech recognition Test Time results - effect of EMG artifact removal}
\end{table*}

\begin{figure}[h]
\begin{center}
\includegraphics[height=4.5cm, width=0.7\linewidth,trim={0.1cm 0.1cm 0.1cm 0.1cm}]{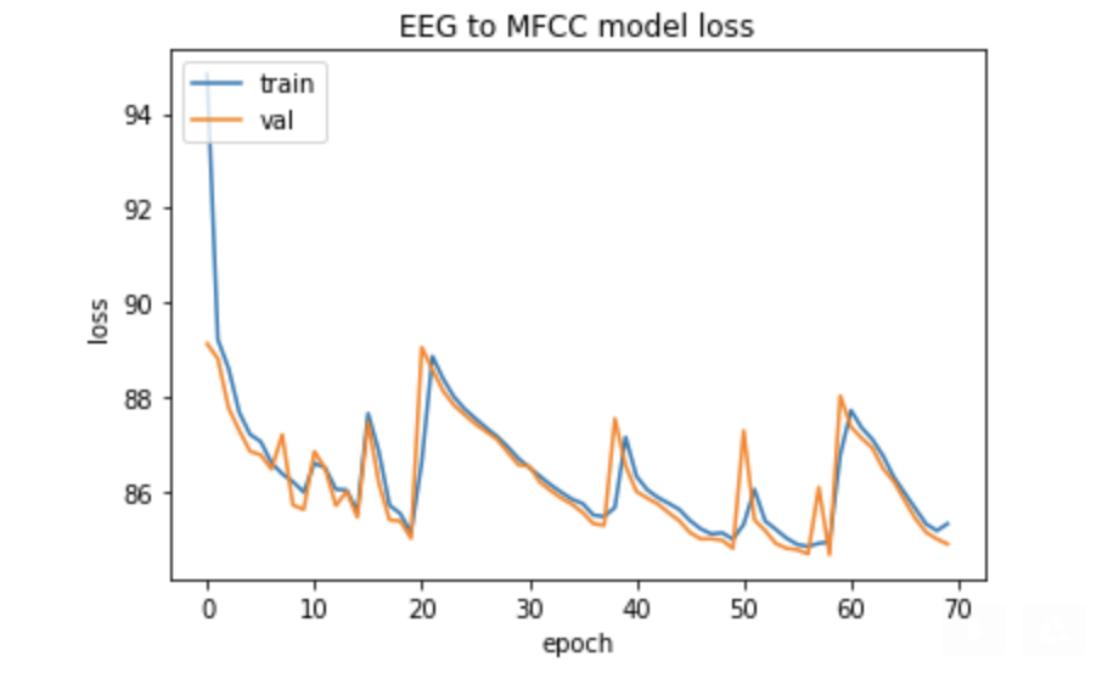}
\caption{Training and validation loss convergence of the regression model} 
\label{1vsall}
\end{center}
\end{figure}

\begin{figure}[h]
\begin{center}
\includegraphics[height=4.5cm, width=0.7\linewidth,trim={0.1cm 0.1cm 0.1cm 0.1cm}]{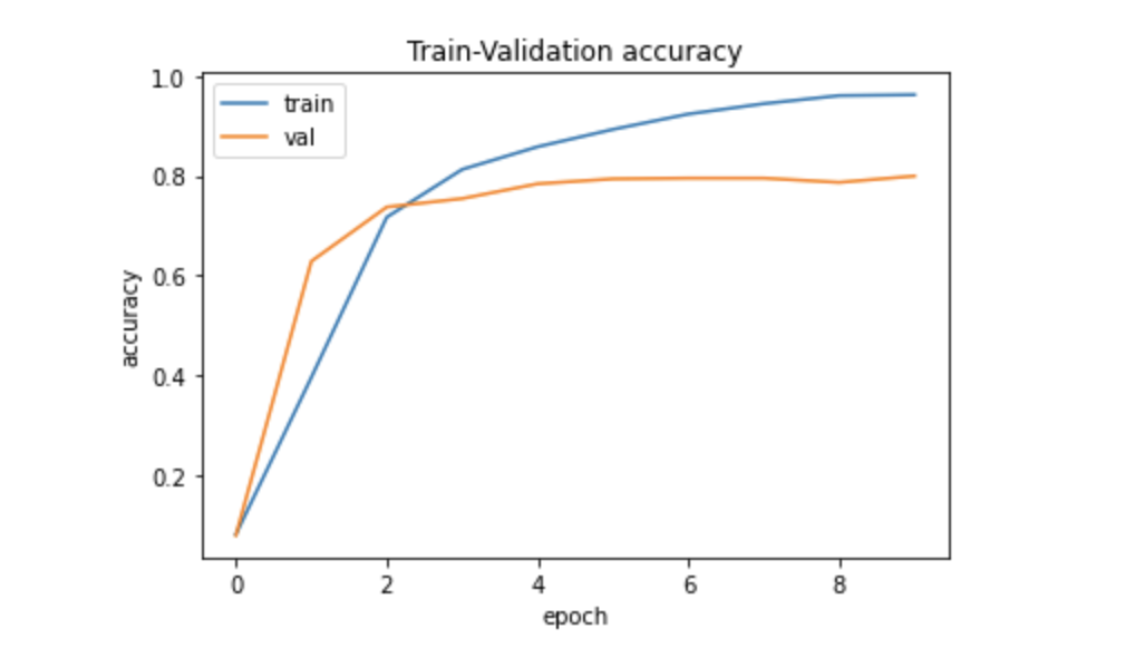}
\caption{Training and validation accuracy of the isolated speech recognition model when trained using MFCC+ acoustic representation in High frequency EEG} 
\label{1vsall}
\end{center}
\end{figure}

\begin{figure}[h]
\begin{center}
\includegraphics[height=7.5cm, width=1.0\linewidth,trim={0.1cm 0.1cm 0.1cm 0.1cm}]{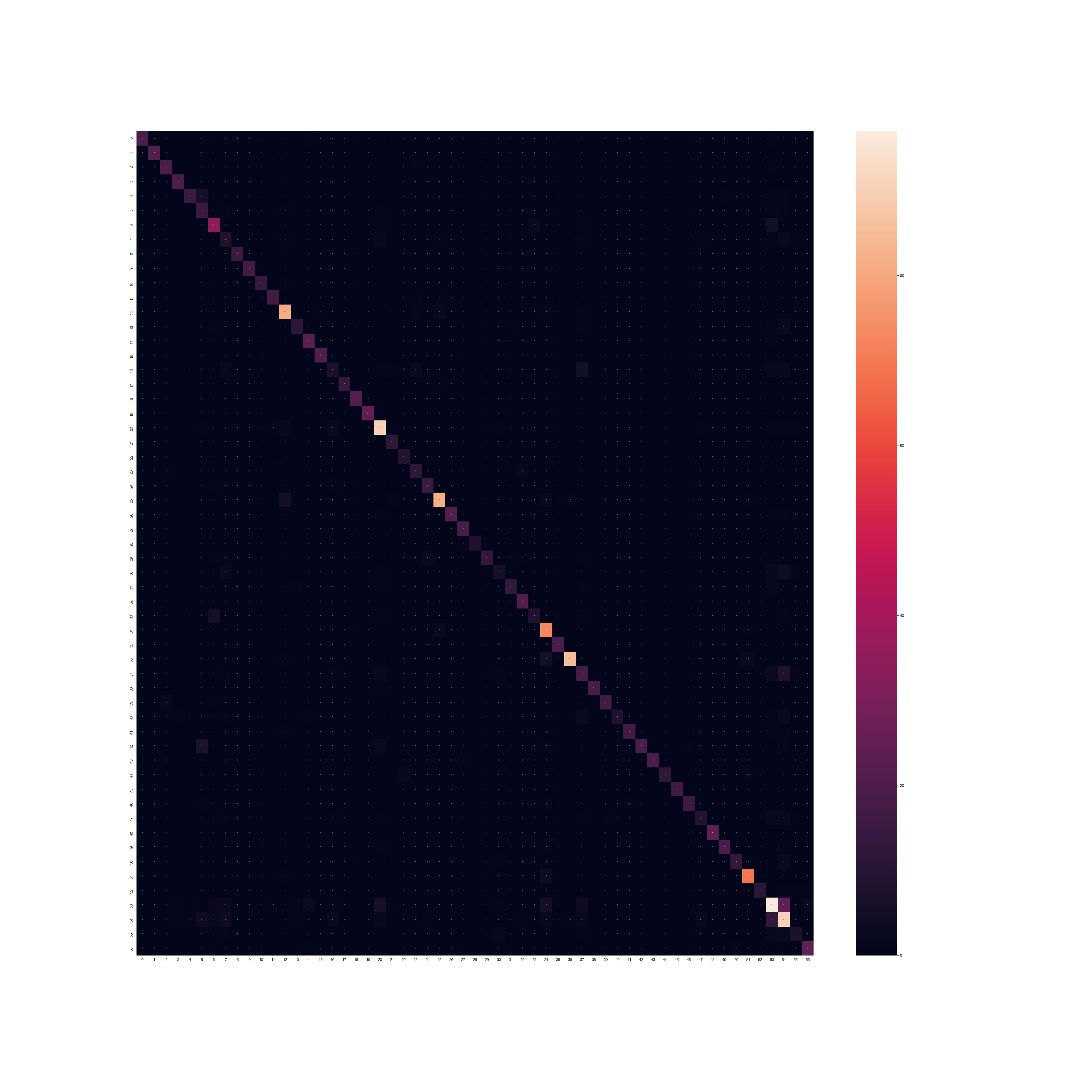}
\caption{Confusion matrix of the isolated speech recognition model during test time when tested using MFCC+ acoustic representation in High frequency EEG} 
\label{1vsall}
\end{center}
\end{figure}

\begin{table*}[!ht]
\tiny
\centering
\begin{tabular}{|l|l|c|c|}
\hline
\textbf{\begin{tabular}[c]{@{}l@{}}Performance\\ Metric(\%)\end{tabular}} & \textbf{MFCC} & \textbf{\begin{tabular}[c]{@{}c@{}}MFCC\\ +\\ Acoustic Representation\\ in High Freq EEG dim 10\end{tabular}} & \textbf{\begin{tabular}[c]{@{}c@{}}MFCC\\ +\\ Acoustic Representation\\ in High Freq EEG dim 145\\ \\ No KPCA\end{tabular}} \\ \hline
\textbf{Accuracy}                                                         & 28.40         & \textbf{81.02}                                                                                        & 28.79                                                                                                               \\ \hline
\textbf{F1 - score}                                                       & 34.93         & \textbf{82.86}                                                                                        & 34.56                                                                                                               \\ \hline
\textbf{precision}                                                        & 75.88         & \textbf{87.23}                                                                                        & 73.66                                                                                                               \\ \hline
\textbf{recall}                                                           & 23.06         & \textbf{79.02}                                                                                        & 22.98                                                                                                               \\ \hline
\end{tabular}
\caption{Isolated Speech recognition test time results-effect of KPCA dimension reduction}
\end{table*}

\begin{table*}[!ht]
\tiny
\centering
\begin{tabular}{|l|l|c|c|c|}
\hline
\textbf{\begin{tabular}[c]{@{}l@{}}Performance\\ Metric(\%)\end{tabular}} & \textbf{MFCC} & \textbf{\begin{tabular}[c]{@{}c@{}}MFCC\\ +\\ Acoustic Representation\\ in High Freq EEG dim 10\\ Temporal Lobe\end{tabular}} & \textbf{\begin{tabular}[c]{@{}c@{}}MFCC\\ +\\ Acoustic Representation\\ in High Freq EEG dim 10\\ Frontal Lobe\end{tabular}} & \textbf{\begin{tabular}[c]{@{}c@{}}MFCC\\ +\\ Acoustic Representation\\ in High Freq EEG dim 20\\ Temporal Lobe and Frontal\\ Lobe\end{tabular}} \\ \hline
\textbf{Accuracy}                                                         & 28.40         & \textbf{81.25}                                                                                                                & 81.14                                                                                                                        & 80.68                                                                                                                                            \\ \hline
\textbf{F1 - score}                                                       & 34.93         & 82.97                                                                                                                         & \textbf{83.78}                                                                                                               & 83.77                                                                                                                                            \\ \hline
\textbf{precision}                                                        & 75.88         & 86.91                                                                                                                         & 89.20                                                                                                                        & \textbf{90.68}                                                                                                                                   \\ \hline
\textbf{recall}                                                           & 23.06         & \textbf{79.53}                                                                                                                & 79.13                                                                                                                        & 78.08                                                                                                                                            \\ \hline
\end{tabular}
\caption{Isolated Speech recognition test time results-effect of EEG sensor reduction}
\end{table*}

\begin{table}[!ht]
\tiny
\centering
\begin{tabular}{|l|l|c|}
\hline
\textbf{\begin{tabular}[c]{@{}l@{}}Performance\\ Metric(\%)\end{tabular}} & \textbf{MFCC} & \textbf{\begin{tabular}[c]{@{}c@{}}MFCC\\ +\\ Acoustic Representation\\ in High Freq EEG dim 10\end{tabular}} \\ \hline
\textbf{Accuracy}                                                         & 78.09         & \textbf{79.84}                                                                                                \\ \hline
\textbf{F1 - score}                                                       & 80.69         & \textbf{82.03}                                                                                                \\ \hline
\textbf{precision}                                                        & 86.02         & \textbf{87.29}                                                                                                \\ \hline
\textbf{recall}                                                           & 76.12         & \textbf{77.46}                                                                                                \\ \hline
\end{tabular}
\caption{Isolated Speech recognition test time results when first half length of the complete speech and EEG signals are used as input}
\end{table}

\begin{table}[!ht]
\tiny
\centering
\begin{tabular}{|l|c|c|c|}
\hline
\textbf{\begin{tabular}[c]{@{}l@{}}Performance \\ Metric(\%)\end{tabular}} & \multicolumn{1}{l|}{\textbf{MFCC}} & \textbf{\begin{tabular}[c]{@{}c@{}}MFCC\\ +\\ Acoustic Representation\\ in High Freq EMG dim 10\end{tabular}} & \textbf{\begin{tabular}[c]{@{}c@{}}MFCC\\ +\\ Acoustic Representation\\ in High Freq EEG dim 10\end{tabular}} \\ \hline
\textbf{Accuracy}                                                          & 28.40                              & 30.15                                                                                                         & \textbf{81.02}                                                                                                \\ \hline
\textbf{F1-score}                                                          & 34.93                              & 37.02                                                                                                         & \textbf{82.86}                                                                                                \\ \hline
\textbf{precision}                                                         & 75.88                              & 80.71                                                                                                         & \textbf{87.23}                                                                                                \\ \hline
\textbf{recall}                                                            & 23.06                              & 24.42                                                                                                         & \textbf{79.02}                                                                                                \\ \hline
\end{tabular}
\caption{Isolated Speech recognition test time results-effect of EMG vs EEG sensors}
\end{table}

\begin{table*}[!ht]
\scriptsize
\centering
\begin{tabular}{|l|l|c|}
\hline
\textbf{Number of sentences} & \textbf{\begin{tabular}[c]{@{}l@{}}MFCC\\ \\ (WER \%)\end{tabular}} & \textbf{\begin{tabular}[c]{@{}c@{}}MFCC\\ +\\ Acoustic Representation\\ in High Freq EEG dim 10\\ \\ (WER \%)\end{tabular}} \\ \hline
1771                         & \multicolumn{1}{c|}{49.84}                                          & \textbf{45.69}                                                                                                              \\ \hline
\end{tabular}
\caption{Continuous Speech Recognition Test time results}
\end{table*}

\section{Conclusion, Limitation and Future work}
In this paper, we proposed a deep learning based algorithm to improve the performance of isolated and continuous speech recognition systems for aphasia, apraxia, and dysarthria speech by utilizing non-invasive neural EEG signals recorded synchronously with the speech. Our proposed algorithm outperformed the baseline results for the task of isolated speech recognition during test time by more than 50\% and at the same time outperforming the baseline results for the more challenging task of continuous speech recognition by a small margin. To the best of our knowledge, this is the first work that demonstrates how to utilize non-invasive neural signals to improve the decoding performance of speech recognition systems for aphasia, apraxia, and dysarthria speech. 
One major limitation of the proposed algorithm is the latency that might be observed when this system is deployed in real-time as the first step is to obtain the acoustic representations in EEG using the trained regression model before it is concatenated with the acoustic features to decode text. All the results presented in this paper are based on the offline analysis. The latency will be a function of the input sequence length, model size, and computational resources (GPU memory and RAM).  
Our future work will focus on validating these results on larger data set as we make progress in our data collection efforts. Future work will also focus on performing more experiments for the task of continuous speech recognition and developing tools to improve the performance of our proposed algorithm. Our aphasia, apraxia, and dysarthria speech-EEG data set will be released to the public to help further advance this interesting and crucial research. 













\bibliographystyle{IEEEtran}

\bibliography{refs}

\begin{thebibliography}{10}
\providecommand{\url}[1]{#1}
\csname url@rmstyle\endcsname
\providecommand{\newblock}{\relax}
\providecommand{\bibinfo}[2]{#2}
\providecommand\BIBentrySTDinterwordspacing{\spaceskip=0pt\relax}
\providecommand\BIBentryALTinterwordstretchfactor{4}
\providecommand\BIBentryALTinterwordspacing{\spaceskip=\fontdimen2\font plus
\BIBentryALTinterwordstretchfactor\fontdimen3\font minus
  \fontdimen4\font\relax}
\providecommand\BIBforeignlanguage[2]{{%
\expandafter\ifx\csname l@#1\endcsname\relax
\typeout{** WARNING: IEEEtran.bst: No hyphenation pattern has been}%
\typeout{** loaded for the language `#1'. Using the pattern for}%
\typeout{** the default language instead.}%
\else
\language=\csname l@#1\endcsname
\fi
#2}}

\bibitem{damasio1992aphasia}
A.~R. Damasio, ``Aphasia,'' \emph{New England Journal of Medicine}, vol. 326,
  no.~8, pp. 531--539, 1992.

\bibitem{benson1996aphasia}
D.~F. Benson and A.~Ardila, \emph{Aphasia: A clinical perspective}.\hskip 1em
  plus 0.5em minus 0.4em\relax Oxford University Press on Demand, 1996.

\bibitem{kent1983acoustic}
R.~D. Kent and J.~C. Rosenbek, ``Acoustic patterns of apraxia of speech,''
  \emph{Journal of Speech, Language, and Hearing Research}, vol.~26, no.~2, pp.
  231--249, 1983.

\bibitem{darley1969differential}
F.~L. Darley, A.~E. Aronson, and J.~R. Brown, ``Differential diagnostic
  patterns of dysarthria,'' \emph{Journal of speech and hearing research},
  vol.~12, no.~2, pp. 246--269, 1969.

\bibitem{krishna2020synthesis}
G.~Krishna, C.~Tran, Y.~Han, M.~Carnahan, and A.~Tewfik, ``Speech synthesis
  using eeg,'' in \emph{Acoustics, Speech and Signal Processing (ICASSP), 2020
  IEEE International Conference on}.\hskip 1em plus 0.5em minus 0.4em\relax
  IEEE, 2020.

\bibitem{anumanchipalli2019speech}
G.~K. Anumanchipalli, J.~Chartier, and E.~F. Chang, ``Speech synthesis from
  neural decoding of spoken sentences,'' \emph{Nature}, vol. 568, no. 7753, p.
  493, 2019.

\bibitem{krishna2019speech}
G.~Krishna, C.~Tran, J.~Yu, and A.~Tewfik, ``Speech recognition with no speech
  or with noisy speech,'' in \emph{Acoustics, Speech and Signal Processing
  (ICASSP), 2019 IEEE International Conference on}.\hskip 1em plus 0.5em minus
  0.4em\relax IEEE, 2019.

\bibitem{fraser2013automatic}
K.~C. Fraser, F.~Rudzicz, N.~Graham, and E.~Rochon, ``Automatic speech
  recognition in the diagnosis of primary progressive aphasia,'' in
  \emph{Proceedings of the fourth workshop on speech and language processing
  for assistive technologies}, 2013, pp. 47--54.

\bibitem{le2016improving}
D.~Le and E.~M. Provost, ``Improving automatic recognition of aphasic speech
  with aphasiabank.'' in \emph{Interspeech}, 2016, pp. 2681--2685.

\bibitem{ballard2019feasibility}
K.~J. Ballard, N.~M. Etter, S.~Shen, P.~Monroe, and C.~Tien~Tan, ``Feasibility
  of automatic speech recognition for providing feedback during tablet-based
  treatment for apraxia of speech plus aphasia,'' \emph{American journal of
  speech-language pathology}, vol.~28, no.~2S, pp. 818--834, 2019.

\bibitem{jacks2019automated}
A.~Jacks, K.~L. Haley, G.~Bishop, and T.~G. Harmon, ``Automated speech
  recognition in adult stroke survivors: Comparing human and computer
  transcriptions,'' \emph{Folia Phoniatrica et Logopaedica}, vol.~71, no. 5-6,
  pp. 286--296, 2019.

\bibitem{green2003automatic}
P.~Green, J.~Carmichael, A.~Hatzis, P.~Enderby, M.~Hawley, and M.~Parker,
  ``Automatic speech recognition with sparse training data for dysarthric
  speakers,'' in \emph{Eighth European conference on speech communication and
  technology}, 2003.

\bibitem{ferrier1995dysarthric}
L.~Ferrier, H.~Shane, H.~Ballard, T.~Carpenter, and A.~Benoit, ``Dysarthric
  speakers' intelligibility and speech characteristics in relation to computer
  speech recognition,'' \emph{Augmentative and Alternative Communication},
  vol.~11, no.~3, pp. 165--175, 1995.

\bibitem{spironelli2009eeg}
C.~Spironelli and A.~Angrilli, ``Eeg delta band as a marker of brain damage in
  aphasic patients after recovery of language,'' \emph{Neuropsychologia},
  vol.~47, no.~4, pp. 988--994, 2009.

\bibitem{hensel2004left}
S.~Hensel, B.~Rockstroh, P.~Berg, T.~Elbert, and P.~W. Sch{\"o}nle,
  ``Left-hemispheric abnormal eeg activity in relation to impairment and
  recovery in aphasic patients,'' \emph{Psychophysiology}, vol.~41, no.~3, pp.
  394--400, 2004.

\bibitem{sarasso2014plastic}
S.~Sarasso, S.~M{\"a}{\"a}tt{\"a}, F.~Ferrarelli, R.~Poryazova, G.~Tononi, and
  S.~L. Small, ``Plastic changes following imitation-based speech and language
  therapy for aphasia: a high-density sleep eeg study,''
  \emph{Neurorehabilitation and neural repair}, vol.~28, no.~2, pp. 129--138,
  2014.

\bibitem{vergin1999generalized}
R.~Vergin, D.~O'Shaughnessy, and A.~Farhat, ``Generalized mel frequency
  cepstral coefficients for large-vocabulary speaker-independent
  continuous-speech recognition,'' \emph{IEEE Transactions on speech and audio
  processing}, vol.~7, no.~5, pp. 525--532, 1999.

\bibitem{chung2014empirical}
J.~Chung, C.~Gulcehre, K.~Cho, and Y.~Bengio, ``Empirical evaluation of gated
  recurrent neural networks on sequence modeling,'' \emph{arXiv preprint
  arXiv:1412.3555}, 2014.

\bibitem{kingma2014adam}
D.~P. Kingma and J.~Ba, ``Adam: A method for stochastic optimization,''
  \emph{arXiv preprint arXiv:1412.6980}, 2014.

\bibitem{srivastava2014dropout}
N.~Srivastava, G.~Hinton, A.~Krizhevsky, I.~Sutskever, and R.~Salakhutdinov,
  ``Dropout: a simple way to prevent neural networks from overfitting,''
  \emph{The journal of machine learning research}, vol.~15, no.~1, pp.
  1929--1958, 2014.

\bibitem{graves2006connectionist}
A.~Graves, S.~Fern{\'a}ndez, F.~Gomez, and J.~Schmidhuber, ``Connectionist
  temporal classification: labelling unsegmented sequence data with recurrent
  neural networks,'' in \emph{Proceedings of the 23rd international conference
  on Machine learning}.\hskip 1em plus 0.5em minus 0.4em\relax ACM, 2006, pp.
  369--376.

\bibitem{graves2014towards}
A.~Graves and N.~Jaitly, ``Towards end-to-end speech recognition with recurrent
  neural networks,'' in \emph{International Conference on Machine Learning},
  2014, pp. 1764--1772.

\bibitem{toshniwal2018comparison}
S.~Toshniwal, A.~Kannan, C.-C. Chiu, Y.~Wu, T.~N. Sainath, and K.~Livescu, ``A
  comparison of techniques for language model integration in encoder-decoder
  speech recognition,'' in \emph{2018 IEEE Spoken Language Technology Workshop
  (SLT)}.\hskip 1em plus 0.5em minus 0.4em\relax IEEE, 2018, pp. 369--375.

\bibitem{krishna20}
G.~Krishna, C.~Tran, M.~Carnahan, and A.~Tewfik, ``Advancing speech recognition
  with no speech or with noisy speech,'' in \emph{2019 27th European Signal
  Processing Conference (EUSIPCO)}.\hskip 1em plus 0.5em minus 0.4em\relax
  IEEE, 2019.

\bibitem{mika1999kernel}
S.~Mika, B.~Sch{\"o}lkopf, A.~J. Smola, K.-R. M{\"u}ller, M.~Scholz, and
  G.~R{\"a}tsch, ``Kernel pca and de-noising in feature spaces,'' in
  \emph{Advances in neural information processing systems}, 1999, pp. 536--542.

\bibitem{goutte2005probabilistic}
C.~Goutte and E.~Gaussier, ``A probabilistic interpretation of precision,
  recall and f-score, with implication for evaluation,'' in \emph{European
  conference on information retrieval}.\hskip 1em plus 0.5em minus 0.4em\relax
  Springer, 2005, pp. 345--359.

\bibitem{flinker2015redefining}
A.~Flinker, A.~Korzeniewska, A.~Y. Shestyuk, P.~J. Franaszczuk, N.~F. Dronkers,
  R.~T. Knight, and N.~E. Crone, ``Redefining the role of broca’s area in
  speech,'' \emph{Proceedings of the National Academy of Sciences}, vol. 112,
  no.~9, pp. 2871--2875, 2015.

\bibitem{chartier2018encoding}
J.~Chartier, G.~K. Anumanchipalli, K.~Johnson, and E.~F. Chang, ``Encoding of
  articulatory kinematic trajectories in human speech sensorimotor cortex,''
  \emph{Neuron}, vol.~98, no.~5, pp. 1042--1054, 2018.

\bibitem{bai2018empirical}
S.~Bai, J.~Z. Kolter, and V.~Koltun, ``An empirical evaluation of generic
  convolutional and recurrent networks for sequence modeling,'' \emph{arXiv
  preprint arXiv:1803.01271}, 2018.

\bibitem{griffiths2019statistical}
P.~Griffiths and J.~Needleman, ``Statistical significance testing and p-values:
  Defending the indefensible? a discussion paper and position statement,''
  \emph{International journal of nursing studies}, vol.~99, p. 103384, 2019.

\end{thebibliography}
\end{document}